\documentclass[twocolumn,pra,aps,showpacs,preprintnumbers,amsmath,amssymb]{revtex4}

\usepackage{graphicx}
\usepackage{dcolumn}
\usepackage{bm}
\usepackage{units}
\usepackage{color}

\newcommand{\be}{\begin{eqnarray}}
\newcommand{\ee}{\end{eqnarray}}
\newcommand{\imagewidth}{0.85\columnwidth}
\newcommand{\ybuion}{\textsuperscript{171}Yb\textsuperscript{+}}

\begin{document}


\title{Two-dimensional cluster-state preparation with linear ion traps}

\author{Harald Wunderlich}
 \email{h.wunderlich@physik.uni-siegen.de}
\author{Christof Wunderlich}
 \email{wunderlich@physik.uni-siegen.de}
\affiliation{Fachbereich Physik, Universit\"at Siegen, 57068 Siegen,
Germany}

\author{Kilian Singer}
\affiliation{Universit\"at Ulm, Institut f\"ur
Quanteninformationsverarbeitung, Albert-Einstein-Allee 11, 89069
Ulm, Germany}

\author{Ferdinand Schmidt-Kaler}
\affiliation{Universit\"at Ulm, Institut f\"ur
Quanteninformationsverarbeitung, Albert-Einstein-Allee 11, 89069
Ulm, Germany}

\date{\today}

\begin{abstract}
We present schemes to prepare two-dimensional cluster states [H. J.
Briegel and R. Raussendorf, Phys. Rev. Lett. {\bf 86}, 910 (2001)]
with atomic ions confined in a microstructured linear ion trap and
coupled by an engineered spin-spin interaction. In particular, we
show how to prepare a $n \times 2$ cluster state by creating a
linear cluster state and adding third-neighbor entanglement using
selective recoupling techniques. The scheme is based on the
capabilities provided by segmented linear Paul traps to confine ions
in local potential wells and to separate and transport ions between
these wells. Furthermore, we consider creating three- and four-qubit cluster
states by engineering the coupling matrix such that through the
periodicity of the time evolution unwanted couplings are canceled.
All entangling operations are achieved by switching of voltages and
currents, and do not require interaction with laser light.
\end{abstract}


\maketitle

\section{Introduction}

Cluster states \cite{Briegel01} are the physical resource needed for
a one-way quantum computer \cite{Raussendorf02-06,Raussendorf03} --
a scheme for measurement-based quantum computation \cite{MBQC07}. It
has been shown that the one-way quantum computer provides a
universal set of quantum gates. Furthermore, cluster states can be
used to efficiently simulate quantum circuits. On the other hand,
some quantum circuits based on the cluster-state model cannot be
interpreted as a network of gates such as the bit-reversal gate
\cite{Raussendorf02-49}.

Not only are cluster states a central ingredient for
measurement-based quantum computing, they are also of interest for
investigating questions of fundamental relevance, for instance,
regarding the robustness of entanglement. The presumption that the
lifetime of entanglement decreases with the number of constituents
of an entangled system and that, therefore, entanglement does not
have to be taken into account when describing the properties of
mesoscopic and macroscopic systems is widespread. This is indeed
true, for example, for Greenberger-Horne-Zeilinger
\cite{Greenberger89} states for which the decoherence rate increases
linearly with the number $N$ of qubits. However, such behavior
does not necessarily hold for other $N$-particle systems. In fact,
many-body states exist where the entanglement between the
constituents is expected to not decay faster than that determined by the
decoherence rate of a single qubit \cite{Duer04}. Cluster states
belong to this class of entangled states.

The topology of cluster states may exist in different dimensions. In
\cite{Nielsen06} it is shown that one-dimensional (1D) cluster states can
be efficiently simulated by classical computers, thus
two-dimensional (2D) cluster states are needed for useful quantum
computation. The first experimental realization of cluster states
was reported in \cite{Mandel03}. In that experiment, the
entanglement was created using controlled collisions between atoms
confined in an optical lattice. Photonic one-way quantum computers
have already been used to implement quantum algorithms, namely,
Grover's algorithm \cite{Walther05} and the Deutsch algorithm on a
four-photon cluster state \cite{Tame07}. Four- and six-qubit
cluster states have also been recently experimentally realized with photons
\cite{Lu07,Park2007}.

Until now, multi-qubit cluster states have not been created with
trapped ions. In \cite{Ivanov08} it was proposed to generate a
linear cluster state of four, five, and six ions in a linear Paul
trap using a gate similar to the M\o{}lmer-S\o{}renson gate
\cite{Molmer99,Sorensen2000}. This would be accomplished by
collective addressing of all ions by means of two laser beams tuned
to the blue- and red-sideband transition of the exploited
vibrational mode. This technique is robust against heating and does
not require the ions to be cooled into their motional ground state.
However, scaling this method to larger numbers of ions becomes
complex.

In this paper we propose schemes to prepare two-dimensional
cluster states with atomic ions confined in a linear
(one-dimensional) trap. In contrast to the scheme described above,
we consider creating cluster states by means of spin-spin coupling
induced by a magnetic-field gradient that creates a state-dependent
force acting on each qubit. Here, laser light is not required to
achieve an entangling gate between the ions. Furthermore, we
consider two-dimensional cluster states, and suggest that our scheme
should be highly scalable. In addition, cooling the ion string to its
motional ground is not necessary.

An easily scalable method for preparing cluster states is applying a
Hamiltonian equivalent to an Ising-type interaction on qubits,
initially prepared in states $|+\rangle = \frac{1}{\sqrt{2}}
(|0\rangle+|1\rangle$) \cite{Briegel01}:
\begin{equation}
\label{eq:HSS}
H = \hbar \sum_{a,a'} J_{a,a'} \frac{1+\sigma_z^{(a)}}{2}
\frac{1-\sigma_z^{(a')}}{2}
\end{equation}
The time evolution $e^{-iHt/\hbar} |+\rangle^{\otimes^n}$ results in
cluster states, if $J_{a,a'} t= \pi + 2k\pi$ with $k \in \mathbb{N}$
and restricting the interaction to next neighbors yields
two-dimensional cluster states. However, in general, experimentally
accessible $J$ couplings $J_{a,a'}$ are of such a form that
preparing cluster states (or performing other operations) is not
trivial.

This paper presents methods to engineer spin-spin couplings
suitable for generating two-dimensional cluster states, a
prerequisite for one-way quantum computing with state-of-the-art
linear ion traps. The outline of the paper is as follows: Sec.
\ref{sec:breaking-up} is a review of how one-way computing with
two-dimensional cluster states can be implemented on a smaller
number of physical qubits by reusing those qubits that have been
measured during the computational process.

In Sec. \ref{sec:spin-coupled} we summarize the relevant
properties of a collection of spin-coupled trapped ions and then
explicitly show in Sec. \ref{sec:Preparing} how to prepare a
$n\times 2$ cluster state with a number of operational steps that is
linear in $n$ using this system.

Section \ref{sec:ions-confined} deals with schemes for cluster state
generation, where the local electrostatic trapping potential
experienced by each ion is individually adjustable. We outline two
schemes that are primarily suited for creating one-dimensional
cluster states, even though they could, in principle, be used to
generate two-dimensional clusters. For implementing the first
scheme, proposed by Mc Hugh and Twamley \cite{McHugh05,McHugh05-71},
the harmonic oscillator frequencies that characterize the local
trapping potentials are set to equal values, resulting in uniform
nearest neighbor couplings. Controlled generation of nearest
neighbor couplings by a choice of appropriate non-uniform trap
frequencies is considered in the second scheme. Then, with the help
of numerical simulations based on an existing microstructured ion
trap, we show that these methods require smaller trap-electrode
structures than currently available to attain coupling constants
that are useful in practice for cluster state generation.

While the above mentioned schemes are based on nearest-neighbor
interactions provided by individual potential wells, Sec.
\ref{sec:Periodic} deals with the question of whether coupling
constants can be engineered in such a way that they fulfill
periodicity conditions imposed on the time-evolution operator
suitable for generating cluster states in one time-evolution step.

In Sec. \ref{sec:including-ion-transport} we propose a scalable
scheme for creating two-dimensional cluster states, which does not
rely on a static placement of the ions in individual wells. Instead
it makes use of the possibilities provided by segmented traps to
adiabatically transport ions, and separate ions held in common
traps, thus allowing us to circumvent problems arising in the
previously mentioned schemes.

The scheme for creating 2D cluster states detailed in Secs.
\ref{sec:ions-confined} and \ref{sec:including-ion-transport} is
underpinned by simulations of electrostatic potentials in a
microstructured ion trap. To be concrete, we used for this purpose
the parameters of a microtrap that is currently being developed.

\subsection{Breaking up a $n \times m$ cluster state into clusters
of size $n \times 2$}
\label{sec:breaking-up}
 In order to make one-way quantum computing
possible first a two-dimensional $n \times m$ cluster state needs to
be prepared and then single-qubit adaptive measurements in
qubit-specific bases are performed in order to achieve a quantum
gate \cite{Raussendorf03}. The basis of a measurement may depend on the
outcome of previous measurements.

The question this paper deals with is: ''How can a two-dimensional
cluster-state be efficiently prepared with trapped ions?''
Generally, preparing a two-dimensional cluster-state turns out to be
difficult for an arbitrary number of qubits. In order to work out an
experimentally feasible procedure for generating cluster states, we
notice that - within the scope of the one-way model - a $n \times m$
cluster state may be broken up into segments of dimension $n\times
2$ as shown by Mc Hugh \cite{McHugh05} and recapitulated for the
reader's convenience in what follows.

The usual way to implement a measurement based quantum computer is to
prepare the entangled state followed by single-qubit measurements.
This procedure is equivalent to the following: imagine the qubits
arranged in a two-dimensional array. Then, the initial state
containing the input data for the quantum gate to be performed is
written upon the first column of qubits. The second column is
prepared in the state $|+\rangle = \frac{1}{\sqrt{2}} (|0\rangle +
|1\rangle)$. After entangling all qubits with their nearest
neighbors via the time evolution generated by Hamiltonian
(\ref{eq:HSS}) with $J_{a,a'}t = \pi$, a $n\times 2$ cluster state
is established. Now, measurements on the first column are performed
in a predetermined basis such that these measurements amount to
executing a desired quantum gate. Due to the entanglement, the
quantum information generated by these measurements is thus
transferred to the second column's qubits. At the same time, the
entanglement between the two columns is erased by the measurement
operation.

That means, the first column may now be used to simulate the third
column of a $n \times m$ cluster. For this purpose, its qubits are
prepared in the state $|+\rangle$ and entangled with the second
column's qubits. Then, single-qubit measurements are performed on
the second column so that the quantum information is transferred to
the first column's qubits. Thus, performing this procedure $m-1$
times, one can simulate a $n\times m$ cluster via a $n\times 2$
cluster.

\subsection{Spin-coupled trapped ions}
\label{sec:spin-coupled}

We consider $N$ ions of mass $m$ confined strongly in two spatial
dimensions ($x$ and $y$ direction) and by a weaker potential in the third
(z direction). This can be achieved, for example, by using a linear
electrodynamic trap where the effective harmonic confinement in the
radial direction, measured in terms of the secular angular frequency
$\nu_r$, is much stronger than in the axial direction characterized
by secular frequency $\nu_z$
\cite{Raizen1992,Schiffer1993,Dubin1993}. The ions are laser cooled
such that their radial degrees of freedom are frozen out and they
form a linear Coulomb crystal.

Each ion provides two internal levels (e.g., hyperfine states)
serving as qubits described by the Pauli operator $\sigma_z$. In
addition, a magnetic field $\vec{B} = (B_0 + bz) \vec{e}_z$ is
applied such that the ions experience a gradient $b$ along the
$z$-axis. Then, the Hamiltonian of the system reads
\cite{Wunderlich02,Wunderlich2003,McHugh05-71}:
 \be \label{eq:H}
H = \frac{\hbar}{2} \sum_{n=1}^{N} \omega_n (z_{0,n})\sigma_{z,n} +
\sum_{n=1}^N \hbar \nu_n a_n^\dagger a_n  \\ \nonumber
 -\frac{\hbar}{2} \sum_{n < m} J_{nm}\sigma_{z,n}\sigma_{z,m}.
 \ee
The first term of the Hamiltonian represents the internal energies
of $N$ qubits where the qubit resonances are given by $\omega_n
(z_{0,n})$, and $z_{0,n}$ is the axial equilibrium position of the
$n$th ion. The second term expresses the collective quantized
vibrational motion in the axial direction of the ions with
eigenfrequency $\nu_n$ of vibrational eigenmode $n$. The last term
describes a pairwise spin-spin coupling between qubits with the
coupling constants.
\begin{equation}
\begin{split}
J_{nm} & = \frac{\hbar}{2 m}\sum_{j=1}^{N} \frac{1}{\nu_j^2}
\frac{\partial \omega_n}{\partial z_n}\vert_{z_{0,n}} \frac{\partial
\omega_m}{\partial z_m}\vert_{z_{0,m}} D_{nj} D_{mj}
\\
& = \frac{\hbar}{2} \frac{\partial \omega_n}{\partial
z_n}\vert_{z_{0,n}} \frac{\partial \omega_m}{\partial
z_m}\vert_{z_{0,m}} (A^{-1})_{nm},
\end{split}
\end{equation}
where $A$ is the Hessian of the trap potential and $D$ is the
unitary transformation matrix that diagonalizes $A$. The eigenvalues
of $A$ are given by $m \nu_j^2$. Therefore, the spin chain can be
interpreted as an $N$-qubit molecule with adjustable coupling
constants $J_{nm}$, an ion spin molecule. If the ions are confined
in a global harmonic potential, then $J_{nm} \propto b^2 /\nu_1^2$.

\begin{figure}[htb]
\begin{center}
\includegraphics[width=300pt]{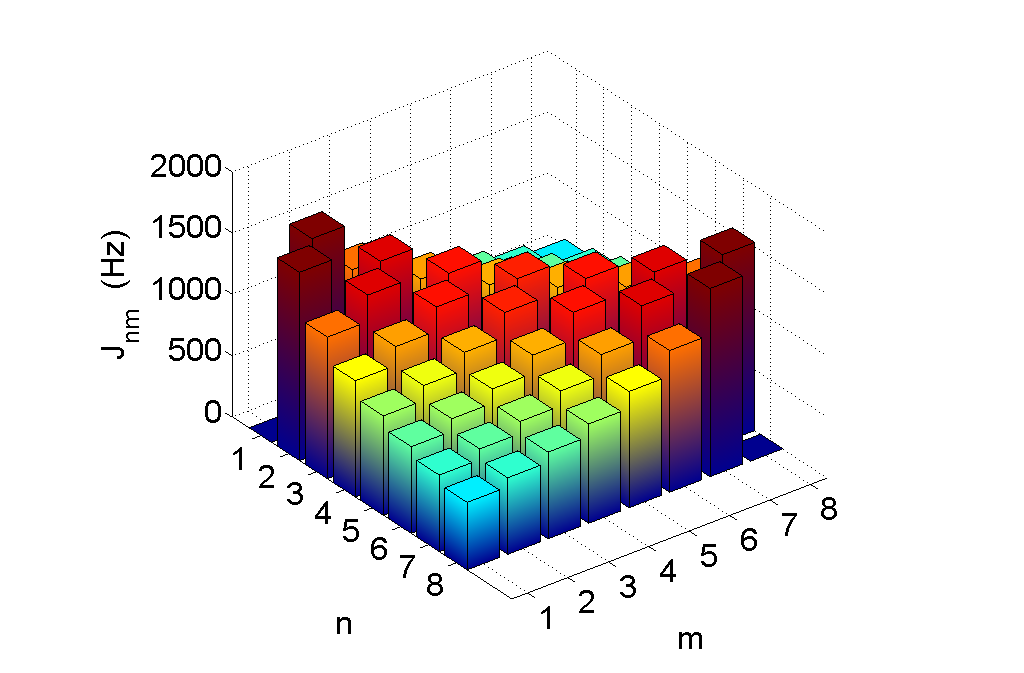}
\caption{(Color online) Coupling constants for eight Yb$^+$ ions in a Paul trap
with axial trap frequency $\nu_1 = 2\pi\times \unit[200]{kHz}$ and a
magnetic field gradient of $\unit[100]{T/m}$.} \label{fig:J8}
\end{center}
\end{figure}

In Fig. \ref{fig:J8}, as a concrete example, the coupling
constants $J_{nm}$ are displayed for eight \ybuion ions in a linear
trap characterized by $\nu_1 = 2\pi\cdot \unit[200]{kHz}$ and a
magnetic field gradient of $\unit[100]{T/m}$.

So far we have a considered a global axial potential confining the
ion string. However, the trapping potential for each ion may be
shaped such that the ions reside in individual harmonic potential
wells. This is accomplished by dividing the dc electrode of a linear
Paul trap into segments to which individual voltages may be applied
that shape the axial potential experienced by the ions. Thus, one or
more ions may be held in local potential wells, and there are
additional handles to change the range and strength of the coupling
constants $J_{nm}$ \cite{McHugh05-71}. Recently, segmented
microstructured traps have been investigated experimentally and
theoretically. Such traps provide the capability of storing ions in
separate potential wells, and of separating and transporting ions
into different trap regions
\cite{Rowe2002,Hensinger2006,Schulz2006,Reichle2006,Hucul2008,Huber08}.

The spin-spin coupling mediated by the vibrational motion in
Eq. (\ref{eq:H}) arises when the ions are exposed to a magnetic
field gradient that induces a state-dependent force. The scheme for
cluster-state preparation proposed here can also be applied to the
case when the required spin-spin coupling is generated by means
other than a magnetic-field gradient. In \cite{Porras2004} it was
shown that an optical state-dependent force may induce a coupling
whose formal description is identical to what is outlined above.
Electrons confined in an array of microstructured Penning traps and
exposed to a spatially varying magnetic field also exhibit a similar
spin-spin coupling \cite{Stahl2005,Ciaramicoli2005,Ciaramicoli2007}.

\section{Preparing cluster states using spin-spin interactions}
 \label{sec:Preparing}
Spin-spin coupling as it appears in Eq. (\ref{eq:H}) may be used
to prepare cluster states. This is achieved in two steps
\cite{Briegel01}: first, all qubits are prepared in the state
$|+\rangle$. Second, the spin-spin coupling according to Eq.
(\ref{eq:H}) is switched on for a time such that $\int J_{nm} dt =
\frac{\pi}{2}+2k\pi$, $k \in \mathbb{N}$ for all qubit pairs $(n,m)$
that are to be entangled. This way of preparing cluster
states provides, in principle, an efficient and scalable method to
generate entangled states.

In actual experiments, the above condition can be fulfilled by
manipulating the coupling constants, by applying a pulse sequence to
selectively realize specific couplings, or by combining these two
methods. In what follows, we will address the issue of creating
suitable interactions for cluster state preparation in detail.

With a string of trapped ions, if all ions are located in the same
harmonic potential well, the $J$ couplings vary in strength
throughout the ion string (see Fig. \ref{fig:J8}). Thus,
achieving controlled dynamics of the system is possible only at high
cost, for example by using selective recoupling pulse sequences
whose length grows quadratically with the qubit number
\cite{Leung00}. Furthermore, the vibrational motion of the ion
string is mediated by the Coulomb repulsion. So $J$ couplings
decrease for non-nearest neighbors. This turns out to be a serious
problem for an efficient preparation of large two-dimensional
cluster-states.

\begin{figure}
\centering
\includegraphics[height=0.2\textheight]{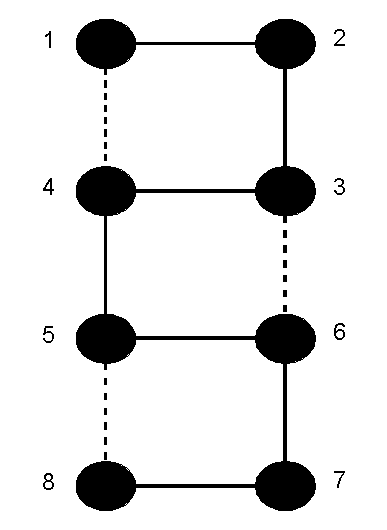}
\caption{Preparing a $4\times 2$ cluster-state: In a first sequence
of operations, a linear 8-qubit cluster-state is created (indicated
by solid lines connecting the qubits). The second sequence then
yields entanglement of third-neighbor qubits (dashed lines)
resulting in the two-dimensional cluster topology.}
\label{fig:cluster8}
\end{figure}

The coarse procedure for creating $n\times 2$ cluster states that
forms the basis for the schemes presented in the remainder of this
paper (unless noted otherwise as in Sec. \ref{sec:Periodic}) is
illustrated in Fig. \ref{fig:cluster8}. First, a linear $n$ qubit
cluster state is prepared. Subsequent entanglement of third-neighbor
qubits then results in the desired two-dimensional cluster.

Alternatively, one could create two four-qubit linear cluster states
(e.g., qubits 1 to 4 and qubits 5 to 8). These linear graph states
could then be converted into box cluster states by local operations
and relabeling of some qubits as described in \cite{Gilbert06}.
Thereafter, entangling qubits 4 and 5 and qubits 3 and 6 would
result in the same graph as depicted in Fig. \ref{fig:cluster8}.

\subsection{Ions confined in individual potential wells}
 \label{sec:ions-confined}
The inhomogeneity of spin-spin couplings in the case of a single
harmonic potential can be treated by locating the ions in the
individual potential wells created in microstructured traps. In
\cite{McHugh05} it is shown that by placing each ion in an
individual potential well, uniform nearest-neighbor interactions
could be achieved. In that scheme, individual harmonic oscillator
potentials (characterized by trap frequencies typically of order
\unit[1]{MHz}) confine a linear array of ions such that they are
spatially separated by $\unit[10]{\mu m}$. The relatively
small distance between neighboring potential wells is necessary to
achieve reasonably large coupling constants $J_{nm}$.

In this arrangement of trap potentials, nearest-neighbor couplings
dominate. Second (third)-neighbor couplings reach values of $\approx
1/6$ ($\approx 1/25$) of the nearest-neighbor couplings (these specific
values result from the choice of the two parameters trap frequency
and ion separation mentioned above). Because of the small
third-neighbor couplings in this arrangement, the general scheme
sketched in Fig. \ref{fig:cluster8} is not well suited for cluster
state generation. In \cite{McHugh05} it was proposed that a
2D cluster state can be created by utilizing nearest- and second-neighbor
couplings in separate steps. This scheme would then require appropriate
refocusing pulse sequences to eliminate undesired couplings (e.g.,
third-neighbor couplings).

An alternative scheme using microstructured traps with electrode
dimensions of order $\unit[10]{\mu m}$ or smaller is sketched in
the following. One may set the potentials of these individual traps
such that, at a given time, only ions $i$ and $i+1$ interact.
This will be the case, if $ \nu_1^j \gg \nu_1^i, \nu_1^{i+1}$ with
$j \neq i,i+1$. This choice of the strength of individual potential
wells ensures strong suppression of non-nearest-neighbor couplings
(below we give a concrete example). Confining the ions in such a
trap configuration and applying a magnetic field gradient for a
suitable time such that $\int J_{i,i+1} dt = \pi /2 $ results in
maximal entanglement between ions $i$ and $i+1$. Thus, a linear
cluster state can be obtained by subsequently performing this
operation on ions $1$ through $N-1$. In order to create a $n \times
2$ cluster state, third-neighbors need to be entangled, that is,
ions 1 and 4, 3 and 6, 5 and 8. This may be accomplished by first
setting $\nu_1^{1}$ through $\nu_{1}^{4}$ to a frequency much lower
than the remaining frequencies to enable coupling between ions 1 and
4. Now, ions 1 and 4 are entangled utilizing spin-spin coupling and
selective refocusing is used on ions 1 through 4 to undo the
unwanted couplings in this quartet of ions, thus realizing the
coupling $J_{14}$. The other third-neighbors are entangled
analogously.

In order to check the feasibility of implementing this scheme with
currently available ion traps we performed numerical calculations of
electrostatic potentials achievable in a typical
microstructured trap.

\subsubsection{Realization with microstructured traps}
 \label{sec:Realization}
The schemes for generating cluster states outlined above require ion
traps with electrode structures at a characteristic length scale of
around \unit[10]{$\mu$m} in order to achieve coupling constants in
the kilohertz regime. Even though such structures appear feasible, they
imply that the distance between ions and a solid-state surface is of
the same order of magnitude, which leads to significant heating
rates of the ions' secular motion
\cite{Turchette2000,Deslauriers2006,Labaziewicz2008}. This, in turn,
is likely to impede precise quantum logic operations
\cite{Wineland1998}. On the other hand, in recent experiments
strongly reduced heating rates have been observed with ion traps in
cryogenic environments \cite{Deslauriers2006,Labaziewicz2008}. Thus,
by sufficiently cooling ion traps, this difficulty that arises with
small electrode structures may be overcome.

\begin{figure}
\centering
\includegraphics[width=\imagewidth]{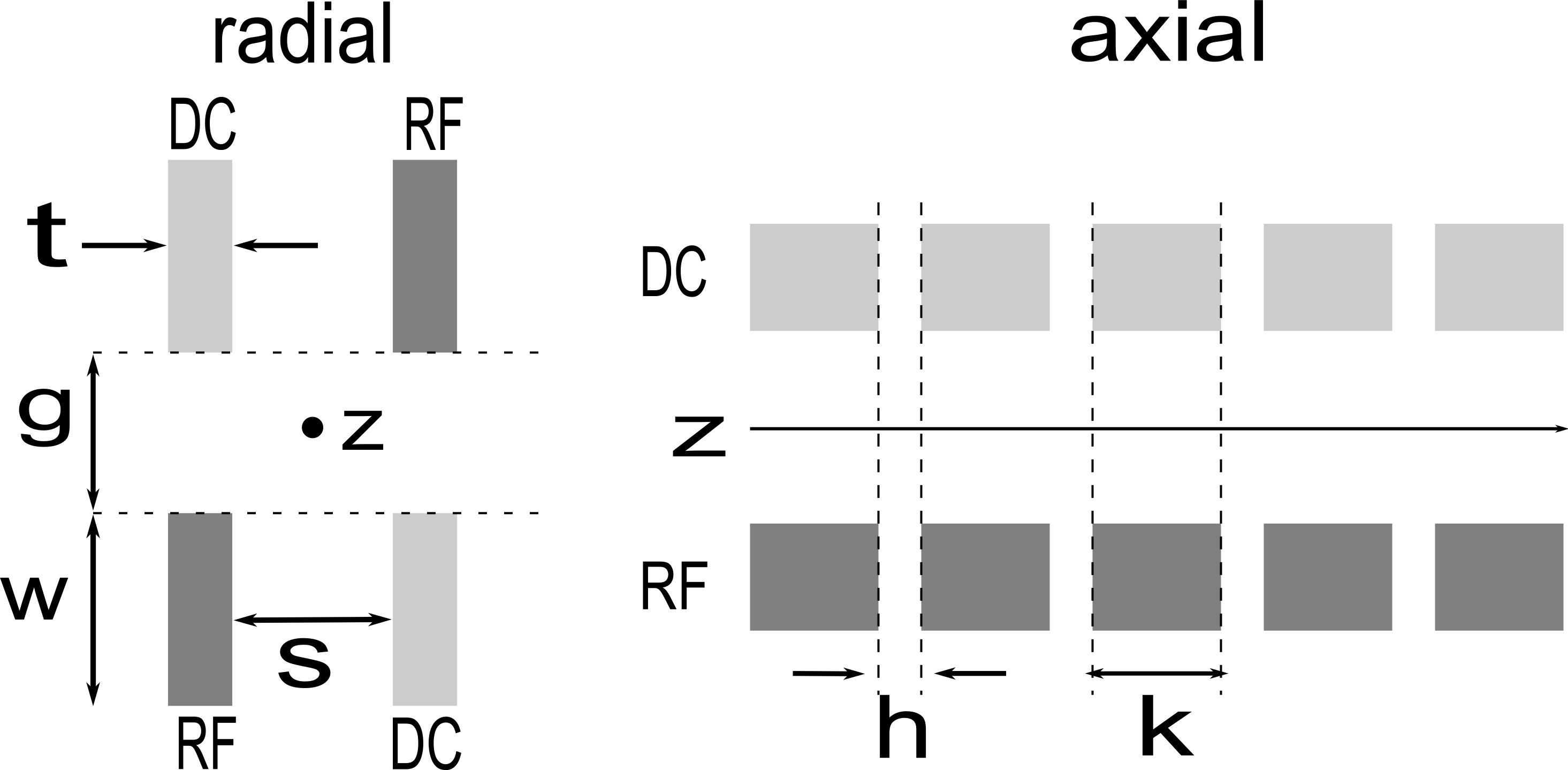}
\caption{(Color online) Sketch of the geometry of a segmented ion
trap considered for potential simulations. The electrode layers are
separated by $s = \unit[350]{\mu m}$, while separation in radial
direction is $g = \unit[250]{\mu m}$. The thickness of the
electrodes is $t=\unit[125]{\mu m}$, the axial length
$k=\unit[100]{\mu m}$, and the axial isolation distance is
$h=\unit[30]{\mu m}$ (compare \cite{Schulz2008}).}
\label{fig:trap-sketch}
\end{figure}

For many existing microstructured traps typical axial lengths of
electrode segments are of order $\unit[100]{\mu m}$ (see Fig.
\ref{fig:trap-sketch}), and isolation spacings between electrode
segments are typically of order $\unit[30]{\mu m}$. Such segmented
microtraps could serve to create individual harmonic oscillator
potentials for each ion. However, with such relatively large trap
structures this would lead to large mutual distances between ions
and thus to small coupling constants $J$. This will make it
difficult to employ the schemes outlined above for efficient cluster
state generation as will be shown now by means of a concrete
example.

For numerical simulations we used the parameters of a microtrap that
is currently being developed. This trap is a three-layer
microstructured segmented trap with two trapping regions. The upper
and lower layers both carry electrodes for applying rf and dc
electric fields. The middle layer serves as a spacer and contains
segments of current-carrying coils that generate a spatially varying
magnetic field \cite{BrueserTBP}. We consider a trapping region in
our potential simulation with the following geometric parameters of
the electrodes: the two electrode layers are separated by the
distance $s = \unit[350]{\mu m}$ and the segmented electrodes are
separated in radial direction by a gap $g = \unit[250]{\mu m}$. The
thickness of the electrodes is $t=\unit[125]{\mu m}$, and the axial
length of each electrode segment amounts to $k=\unit[100]{\mu m}$.
For isolation, the electrode segments are divided by a gap of
$h=\unit[30]{\mu m}$.

We consider 17 segment electrode pairs for the potential simulation.
The simulation was created using the \textit{isim} package \cite{Schulz2008}, which is based on boundary-element methods. All
coupling constants calculated in this section are done for six
\ybuion ions.

This ion trap is structured such that for each ion an individual
potential well may be applied and thus the trap frequencies can be
individually set. It is then possible to generate a sequence of
single nearest-neighbor couplings with suppressed non-nearest-
neighbor couplings. As noted above, setting two neighboring
oscillators' frequencies to small values compared to all other's
allows selective coupling of a single pair of qubits.

\begin{figure}[h]
\begin{center}
    \includegraphics[width=260pt]{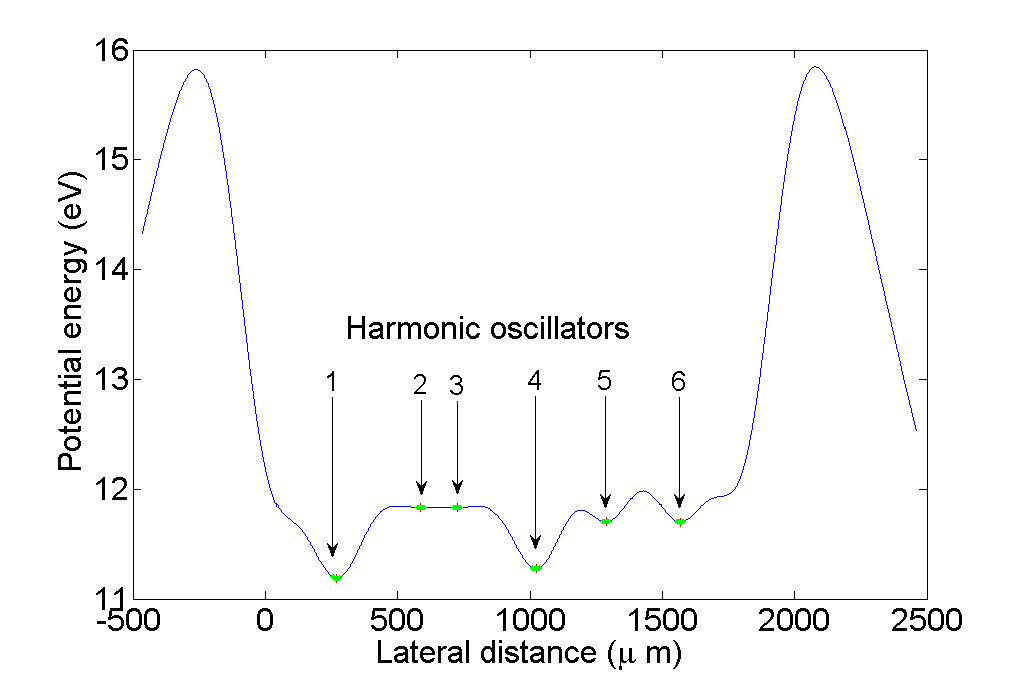}
    \caption{(Color online) Potential simulation for the voltage configuration shown
    in Table \ref{tab:VS23}. The arrows indicate local potential minima (i.e.,
    the equilibrium position of the respective ion).
    Oscillators 2 and 3 have small frequencies
    compared to the other oscillators so that $J_{23}$ is the
    dominating coupling.}
    \label{fig:trap_pot23}
  \end{center}
\end{figure}

Coupling of, for example, only ions number 2 and 3 can be attained
by applying the voltage configuration given in Table \ref{tab:VS23}
to the trap dc electrodes. These voltages result in the simulated
potential shown in Fig. \ref{fig:trap_pot23}. A polynomial fit up to
second order yields trap frequencies for oscillators 2 and 3 of
$\unit[0.35]{MHz}$ and $\unit[0.27]{MHz}$, whereas all other
frequencies are between $0.8$ and $\unit[1.6]{MHz}$.
Furthermore, the distance between oscillators 2 and 3 is smaller by
a factor of around 2 compared to the other oscillator distances.
The nearest-neighbor coupling constants shown in Table \ref{tab:J23}
illustrate that $J_{23}$ dominates over all other couplings by 2 orders
of magnitude as desired. But due to the large distance between ions
of $\unit[140]{\mu m}$, even this dominating coupling is very small.
So segmented microtraps with much smaller axial electrode lengths
and isolation spacings (of order $\unit[10]{\mu m}$) would be
required to achieve  $J$ couplings in the kilohertz range.

\begin{table}

\begin{ruledtabular}
\begin{tabular}{|c|c|c|c|c|c|c|c|c|c|}
Segment No. & 1 & 2 & 3 & 4 & 5 & 6 & 7 & 8 & 9
\\
Voltage / V & 48 & -8 & 48 & 0  & 37.1 & 18.3 & 27.4 & 18.3 & 36.8
\\
\hline Segment No.  & 10 & 11 & 12 & 13 & 14 & 15 & 16 & 17 &
\\
Voltage / V  & 0 & 48 & 0 & 48 & 0 & 48 & -8 & 48 &
\end{tabular}
\end{ruledtabular}
\normalsize \caption{Example for voltage configuration for the
microstructured trap described in the text to establish the single
coupling $J_{23}$ in a string of six \ybuion ions while strongly
suppressing all other couplings.} \label{tab:VS23}
\end{table}

\begin{table}
\begin{tabular}{|c|c|c|c|}
\hline $i$ & $d_{i,i+1}/\unit{\mu m}$ & $\nu_i/\unit{MHz}$ &
$J_{i,i+1}/\unit{Hz}$
\\
\hline
1 & 320 & 1.65 & 0.001
\\
2 & 138 & 0.35 & 0.610
\\
3 & 297 & 0.27 & 0.006
\\
4 & 266 & 1.16 & 0.001
\\
5 & 279 & 0.83 & 0.001
\\
6 & - & 0.98 & -
\\
\hline
\end{tabular}
\caption{Distances $d_{i,i+1}$ between the individual potential
wells $i/i+1$ and axial trap frequencies $\nu_i$ determined from
polynomial fits up to second order around the potential minima shown
in Fig. \ref{fig:trap_pot23}. These parameters clearly lead to the
domination of the coupling $J_{23}$ by two orders of magnitude over
all other couplings. In this way controlled nearest-neighbor
couplings may be implemented without the need for refocusing
pulses. To increase $J$-couplings smaller distances between ions,
and therefore smaller electrode structures are required}
\label{tab:J23}
\end{table}

\subsubsection{Periodicity of the time-evolution operator}
 \label{sec:Periodic}
The schemes outlined above for preparing a two-dimensional
cluster-state are based on the generation of a linear cluster-state
of ions and subsequent third-neighbor couplings. These
third-neighbor couplings, while undoing the unwanted next-neighbor
(NN) couplings, could be accomplished by selective recoupling
techniques (compare Sec. \ref{sec:including-ion-transport}).
However, simultaneous coupling of all qubit pairs would be
advantageous. In this section we show that tailoring the
time-evolution operator, that is, imposing a suitable periodicity
condition by sculpting the J couplings, allows for creation of a
cluster state for three qubits, which corresponds to a triangle
graph, and the creation of a linear cluster state for four qubits in
one time-evolution step.

The general form of the time-evolution operator is:
\begin{equation}
\label{eq:Uperiodic}
U = \prod_{i < j} \exp(i \Theta_{ij} \sigma_{z,i}\sigma_{z,j}) \ ,
\end{equation}
where $\Theta_{ij} = \int J_{ij} dt$. In order to obtain
cluster states by spin-spin coupling, the $\Theta_{ij}$ need to take
on values of $\frac{\pi}{4}, \frac{\pi}{4}+2\pi,
\frac{\pi}{4}+4\pi,...$, whereas $\Theta$ values of $0, 2\pi, 4\pi,
...$ transform the time-development operator into the identity
operator. So using periodicity to entangle three qubits requires
the following $J$ matrix:
\begin{equation}
\Theta = \int J dt =
\begin{pmatrix}
                                    &   \frac{\pi}{4}+2 k \pi & \frac{\pi}{4} \\
\frac{\pi}{4}+2 k \pi   &    & \frac{\pi}{4}+2 k \pi \\
\frac{\pi}{4}                   & \frac{\pi}{4}+2 k \pi  &
\end{pmatrix}
\end{equation}
with $k \in \mathbb{N}$. Applying this time evolution to qubits
prepared in $|+\rangle^{\otimes^3}$ results in the three-qubit cluster
state
 \be
\begin{split}
|C_3\rangle & = \exp(i \frac{\pi}{4} (\sigma_{z,1} \sigma_{z,2}+\sigma_{z,1}\sigma_{z,3}+\sigma_{z,2}\sigma_{z,3} ))|+\rangle^{\otimes^3}
\\
& =_{l.u.} \frac{1}{2}(|000\rangle+|111\rangle) \ .
\end{split}
 \ee
Here l.u. denotes equivalence up to local unitaries.

One may also utilize the periodicity  relation to realize only
NN interactions. In this case, the $J$ matrix reads:
\begin{equation}
\Theta = \int J dt =
\begin{pmatrix}
                                &  \frac{\pi}{4} + 2 k_3 \pi & 2 k_2 \pi    & 2 k_1 \pi \\
\frac{\pi}{4} + 2 k_3 \pi &                                     &   \frac{\pi}{4} + 2 k_4 \pi           &  2 k_2 \pi  \\
2 k_2 \pi                       & \frac{\pi}{4} + 2 k_4 \pi  &              & \frac{\pi}{4} + 2 k_3 \pi \\
2 k_1 \pi                       & 2 k_2 \pi
& \frac{\pi}{4} + 2 k_3 \pi &
\end{pmatrix}
\label{eq:Theta}
\end{equation}
with  $k_1, k_2, k_3, k_4 \in \mathbb{N}$. The time evolution
(\ref{eq:Uperiodic}) describes an effective nearest-neighbor
interaction, thus generating the linear four-qubit cluster state:
 \be
\begin{split}
|C_4\rangle = & \exp(i \frac{\pi}{4} (\sigma_{z,1} \sigma_{z,2}+\sigma_{z,2} \sigma_{z,3}+\sigma_{z,3} \sigma_{z,4})) |+\rangle^{\otimes^4}
\\
& =_{l.u.} \frac{1}{2} (|0000\rangle + |0011\rangle+|1100\rangle -
|1111\rangle) \ .
\end{split}
 \ee
Replacing the matrix elements $\Theta_{14} = 2k_1\pi = \Theta_{41}$
in matrix \ref{eq:Theta} by $\Theta_{14} = 2k_1\pi + \pi/4 =
\Theta_{41}$ would allow for generating a 2D cluster state.

\begin{figure}[htb]
\begin{center}
      \includegraphics[width=\imagewidth]{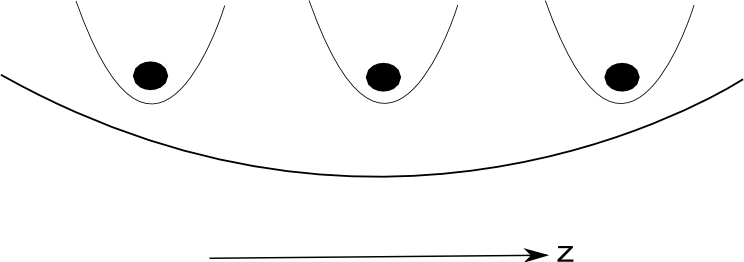}
    \caption{Locating ions in individual microtraps and superposing
     a harmonic long-range potential serves to create coupling constants
     suitable for periodic entanglement.}
    \label{fig:segtrap}
  \end{center}
\end{figure}

In the following we show how coupling matrices can be achieved that
fulfill the desired periodicity. As before (Sec.
\ref{sec:Realization}), the ions are placed in individual harmonic
oscillator potential wells with adjustable frequencies. Superposing
a long range harmonic potential affecting all ions, one more degree
of freedom is available (sketched in Fig. \ref{fig:segtrap}). So the
problem reduces to finding trap frequencies resulting in $J$ couplings
that fulfill the periodicity relation. The coupling constants are
functions of the ions' equilibrium positions, which can be
calculated analytically only for two and three ions. Furthermore, a change
in only one trap parameter, say in one trap frequency, affects all
couplings simultaneously. So due to the highly non-linear nature of
the coupling constants as well as the sensitivity to parameter
alterations, finding trap configurations that are suitable for
utilizing the periodicity condition becomes increasingly difficult
with the number of ions involved.  Here we present empirically found
parameters.

\begin{figure}[htb!]
    \includegraphics[width=300pt]{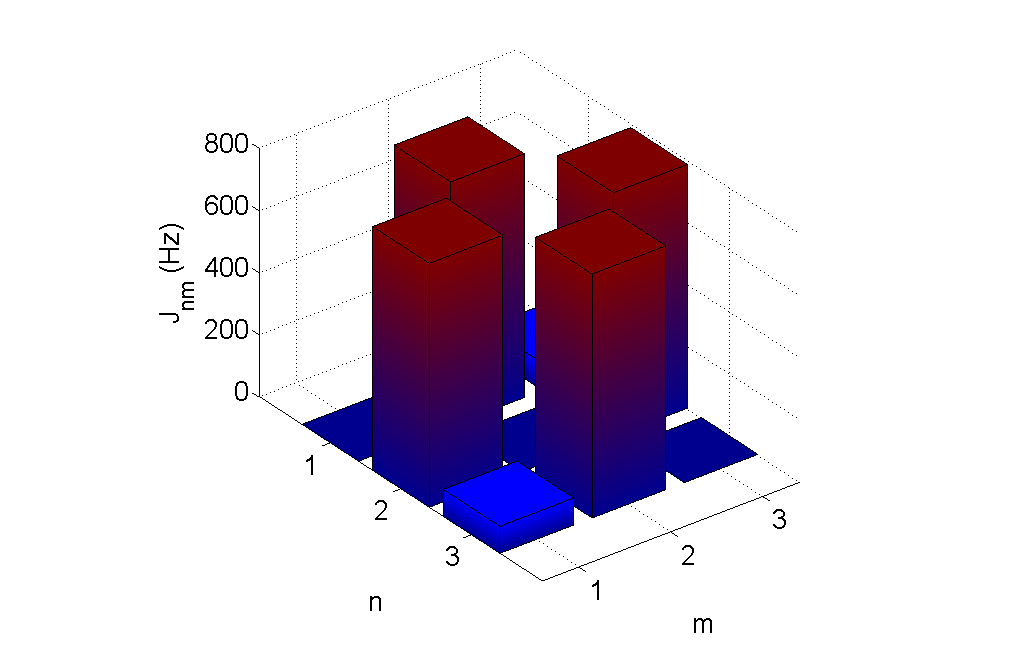}
    \caption{(Color online) Coupling constants suitable for generating a triangle graph state
    in one time evolution for three Yb$^+$ ions in individual
    potential wells superimposed on a harmonic long-range potential of
    frequency $\omega = 2\pi\cdot 100~\unit{kHz}$. Trap frequencies of
    traps 1 and 3 are $2\pi\cdot 277~\unit{kHz}$. The middle
    individual trap has frequency $2\pi \cdot 100~\unit{kHz}$.}
    \label{fig:J3adj}
\end{figure}

Consider three Yb$^+$ ions in individual potential wells superposed
by a harmonic long-range potential of frequency $\omega = 2\pi\times
\unit[100]{kHz}$. Trap frequencies of traps 1 and 3 are $2\pi\times
\unit[277]{kHz}$. The middle individual trap has frequency $2\pi
\times \unit[100]{kHz}$ (see Figs. \ref{fig:segtrap} and
\ref{fig:J3adj}). The potential wells are separated by a distance of
$\unit[20]{\mu m}$, and a magnetic field gradient of
$\unit[100]{T/m}$ is applied along the trap axis. These parameters
result in the following $J$ matrix:
\begin{equation}
\label{eq:J3adj} J / (100~\unit{Hz}) =
\begin{pmatrix}
               &   7.85  & 0.87 \\
   7.85     &   &  7.85 \\
   0.87 &  7.85           &         \\
\end{pmatrix}
\end{equation}
These couplings are useful for cluster state preparation since the
periodicity relation is fulfilled:
\begin{equation}
\frac{J_{21}}{J_{31}} = 9.02 \approx \frac{2\pi+\pi/4}{\pi/4}
\end{equation}

In an analogous fashion, we present parameters  useful for generating a
four-qubit NN interaction. The four ions are again located in
individual traps of frequency $\omega_1 = \omega_4 = 2\pi\times 415
~\unit{kHz}$ and $\omega_2 = \omega_3 = 2\pi \times 280
~\unit{kHz}$. A harmonic long-range potential of frequency $\omega =
2\pi\times 239~ \unit{kHz}$ is then applied. The individual traps
are separated by $\unit[5]{\mu m}$, and the magnetic-field gradient
is again $\frac{\partial B}{\partial z} = 100 ~\unit{T/m}$. With
these parameters, we obtain the following $J$ matrix:
\begin{equation}
\label{eq:J4adj} J / (100~\unit{Hz}) =
\begin{pmatrix}
           &  4.33 &  2.08  &  1.05 \\
   4.33 &        &  4.36 &   2.08 \\
   2.08 &   4.36 &     &  4.33 \\
   1.05 &  2.08  &  4.33 &
\end{pmatrix}
\end{equation}
which results in following periodicity relations:
\begin{equation}
\begin{split}
\frac{J_{32}}{J_{41}} &= 4.15 \approx \frac{4\cdot 2\pi+ \pi/4}{2\pi} \\
\frac{J_{21}}{J_{41}} &=  4.12 \approx \frac{4\cdot 2\pi+ \pi/4}{2\pi} \\
\frac{J_{31}}{J_{41}} &= 1.98 \approx \frac{2\cdot 2\pi}{2\pi}.
\end{split}
\end{equation}

We conclude from the case of three ions that  periodic entanglement
is in principal possible, but as can be seen from the case of four
ions, finding appropriate trap frequencies becomes more and more
difficult with higher number of ions due to the non-linearity of the
problem. We conjecture that numerical approaches such as genetic
algorithms could be efficiently applied to find optimal parameter
configurations.

In Sec. \ref{sec:Realization} it was demonstrated that suitable
voltage configurations can be found to implement a desired coupling
matrix. In order to attain reasonably large coupling constants small
electrode structures (of order \unit[10]{$\mu$m}) are required.
For the scheme presented in Sec. \ref{sec:Periodic},
relying on appropriate periodicity of the time-evolution
operator, again small electrode structures are required to be
implemented efficiently.

\subsection{Including ion transport for generating cluster states}
\label{sec:including-ion-transport} Now we turn to the description
of a different scheme for generating cluster states, which takes
advantage of the capabilities of segmented ion traps, even with
relatively large electrodes. Segmented ion traps may not only serve
for creating an array of linearly arranged potential wells, but also
for transporting ions into different trap regions as well as for
separation of ions held in a common trap into two distinct traps
\cite{Rowe2002,Hensinger2006,Schulz2006,Reichle2006,Hucul2008,Huber08}.

Adiabatic ion transport using segmented microtraps has been
demonstrated by Rowe et al. \cite{Rowe2002}. In this experiment a $
^9$Be$^+$ ion was transferred between trap locations 1.2 mm apart in
\unit[50]{$\mu$s} with almost unit efficiency. Furthermore,
separation of two ions held initially in a common trap into distinct
traps was demonstrated. This was accomplished by using a five
electrode configuration. Fast non-adiabatic transport of
\textsuperscript{40}Ca\textsuperscript{+} ions was reported in
\cite{Huber08}. The experimental results show a success rate of
99.0(1)\% for a transport distance of $2 \times 2 \unit{mm}$ in a
round-trip time of $T = \unit[20]{\mu s}$. Application of optimal
control theory is planned to achieve lower excitation of vibrational
motion in the future.

A $n\times 2$ cluster is prepared using the scheme illustrated in
Fig. \ref{fig:cluster8}. This generation of cluster states is
accomplished in two operational sequences. During the first
sequence nearest-neighbor couplings are established that lead to the
preparation of a one-dimensional cluster state. The second sequence
then establishes couplings between third-neighbor qubits and serves
to create a 2D cluster state of eight ions. These two sequences
will be detailed in what follows. First, we outline the sequences
and state the required time evolution. Then, numerical simulations
of the potentials of a microstructured ion trap, which is currently
being tested, will serve to illustrate the feasibility of the
proposed scheme.

\subsubsection{First sequence: creating a 1D cluster state}
The first sequence itself consists of two steps. During the first
time step with duration $t_1$, pairs of ions, namely ions number 1
and 2, 3 and 4, 5 and 6, 7 and 8 occupy a common trap potential.
Switching on the magnetic field gradient for a desired time results
in $N/2$ uniform nearest-neighbor couplings:
 \be
 H_{t_1} = -\frac{\hbar}{2} J_{1,2}(\sigma_{z,1}\sigma_{z,2}+\sigma_{z,3}\sigma_{z,4}+\sigma_{z,5}\sigma_{z,6}+\sigma_{z,7}\sigma_{z,8})
 \ee
The duration $t_1$ is chosen such that $\int_0^{t_1} J dt =
\frac{\pi}{2}$ is fulfilled and the time-evolution operator
 \be
 \label{eq:Ut1}
 U_{t_1} = \exp(i \frac{\pi}{4} (\sigma_{z,1}\sigma_{z,2}+\sigma_{z,3}\sigma_{z,4}+\sigma_{z,5}\sigma_{z,6}+\sigma_{z,7}\sigma_{z,8}))
 \ee
is obtained.

At the end of the first time interval, the magnetic-field gradient
is turned off (i.e., the spin-spin is zero), and the ions sharing a
common trap are separated and transported into potential wells such
that ions 2 and 3, 4 and 5, 6 and 7 occupy a common trap. When the
ion transport is finished, the magnetic-field gradient is switched
on again during time $t_2$, thus resulting in the other half of
NN couplings:
 \be
 \label{eq:Ut2}
 U_{t_2} = \exp(i \frac{\pi}{4} (\sigma_{z,2}\sigma_{z,3}+\sigma_{z,4}\sigma_{z,5}+\sigma_{z,6}\sigma_{z,7}))
 \ee

\begin{figure}[htb]
    \includegraphics[width=260pt]{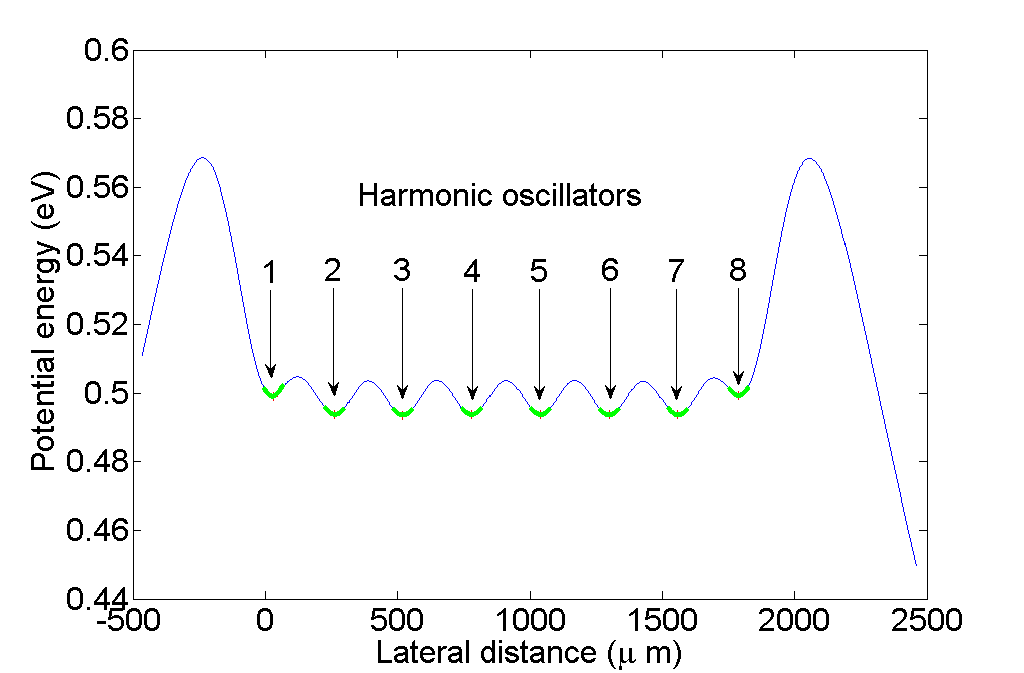}
    \caption{(Color online) Simulated potential for the voltage configuration shown
    in Tab.\ref{tab:V200k}. Six of eight potential wells have uniform
    axial frequencies of $\unit[200]{kHz}$ as well as equidistant
    axial spacings of $\unit[260]{\mu m}$.}
    \label{fig:trap_pot200}
\end{figure}

The time evolution during the first sequence described above (i.e.,
between $t=0$ and $t=t_2$) requires $N/2$ potential wells of equal
axial trap frequency for $N$ ions. We now show that the required
axial potential can be realized using the segmented microtrap
introduced in Sec. \ref{sec:spin-coupled}. Applying a voltage of
$\unit[1.6]{V}$ to the outermost electrodes, and an alternating
series of $\unit[2]{V}$ and $\unit[0]{V}$, respectively, on the
remaining dc electrode pairs (Table \ref{tab:V200k}), results in
harmonic axial potentials of frequency $\omega = 2\pi\times
\unit[200]{kHz}$ (compare Fig. \ref{fig:trap_pot200})

\begin{table}
\begin{ruledtabular}
\begin{tabular}{|c|c|c|c|c|c|c|c|c|c|c|c|c|c|c|c|c|}
1.6 & 0 & 2 & 0 & 2 & 0 & 2 & 0 & 2 & 0 & 2 & 0 & 2 & 0 & 2 & 0 &
1.6
\\
\hline
\end{tabular}
\end{ruledtabular}
\caption{Voltage configuration for creating 8 harmonic potential
wells such that oscillators 2 to 7 have uniform axial frequencies of
$200~\unit{kHz}$. Voltage at electrode pairs 1 and 17 is lower in
order to reduce fringe effects.} \label{tab:V200k}
\end{table}

Six of the eight created harmonic oscillators have the same
frequency, so that the scheme for preparing a two-dimensional
eight-qubit cluster state can be realized with this axial potential.
Keeping the pairs of ions 1 and 2 through 7 and 8 in potential wells
2 through 5 for the time $t_1$ results in the time evolution
according to Eq. (\ref{eq:Ut1}) when the magnetic-field gradient
is switched on. Thereafter, the field gradient is deactivated, and
the ions are rearranged, such that ions 1 and 8 occupy potential
wells 2 and 6 respectively, and the pairs of ions 2 and 3 through 6
and 7 are located in wells 3 through 5. Reactivating the field
gradient for a time $t_2$ leads to the evolution given in Eq.
(\ref{eq:Ut2}), thus creating an eight-qubit linear cluster state.

The distances between the minima of the potential wells are shown in
Table \ref{tab:dist200k}. The characteristic value of the distances
between the traps 2 to 7 is simply given by $2\cdot(k+h) =
\unit[260]{\mu m}$. Deviating values of oscillators 1 and 8 can be
explained by fringe effects.

\begin{table}
\caption{Distances between the minima of the potential wells in
$\unit{\mu m}$ for the voltage configuration shown in Tab.
\ref{tab:V200k}} \label{tab:dist200k}
\begin{ruledtabular}
\begin{tabular}{ccccccc}
230   &    260  &     260  &     260  &     260 & 259 & 231
\end{tabular}
\end{ruledtabular}
\end{table}

\begin{figure}[htb]
\begin{center}
\includegraphics[width=300pt]{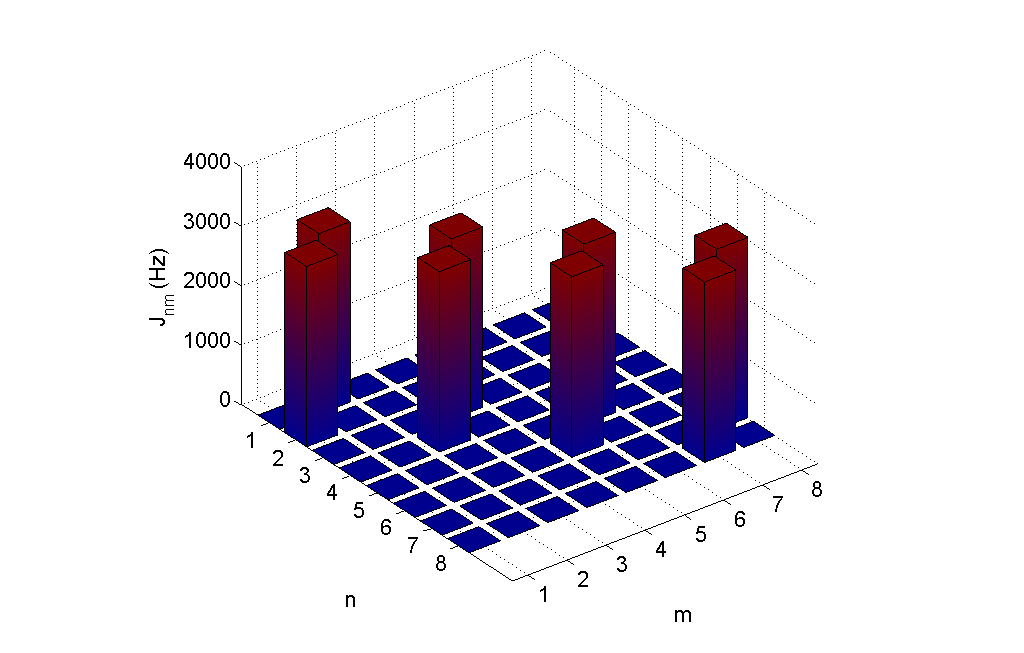}
\caption{(Color online) Coupling constants for 8 ions stored in individual harmonic
potential wells (shown in Fig. \ref{fig:trap_pot200}). Each
potential well is occupied by two ions, yielding uniform NN
couplings of $\unit[3]{kHz}$ between ions number 1 and 2, 3 and 4,
etc.. These coupling constants are required for the first time
evolution in the first sequence of operations at the end of which a
2D cluster state is obtained. Non-nearest-neighbor couplings are
suppressed by four orders of magnitude.} \label{fig:NNhalf1}
\end{center}
\end{figure}

\begin{figure}[htb]
\begin{center}
\includegraphics[width=300pt]{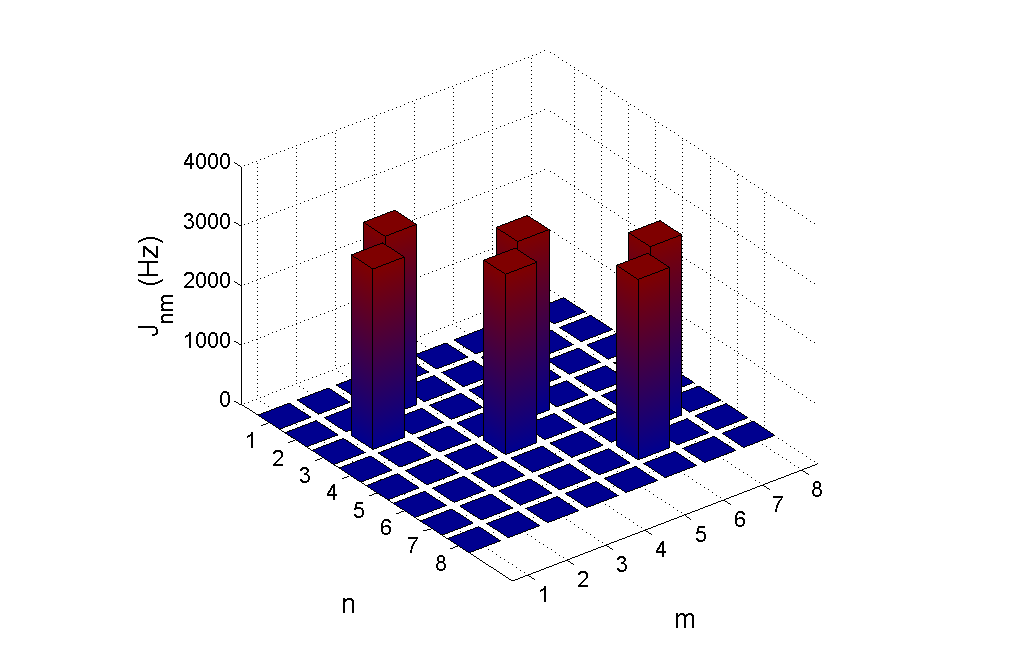}
\caption{(Color online) After creating the NN couplings shown in Figure \ref{fig:NNhalf1},
 the magnetic field gradient is switched off. Then, ions sharing a common
 oscillator are separated and transported into four other potential wells,
 so that ions 2 and 3, 4 and 5 etc. occupy now a common well.
 Switching on a magnet field gradient of $\unit[100]{T/m}$ results
 in uniform NN couplings of $\unit[3]{kHz}$ again.
 This completes the second step of the first sequence of operations required
 for a 2D cluster.}
\label{fig:NNhalf2}
\end{center}
\end{figure}

Trap frequencies of $\omega= 2\pi\times \unit[200]{kHz}$ and a
magnetic-field gradient of $\partial_z B = \unit[100]{T/m}$ result
in coupling constants of $\approx \unit[3]{kHz}$. In Fig.
\ref{fig:NNhalf1} the coupling constants for the time evolution for
$0< t < t_1$ are displayed and Fig. \ref{fig:NNhalf2} shows the
coupling values for $t_1< t < t_2$. Non-nearest-neighbor couplings
are suppressed by 4 orders of magnitude due to the large distance of more than $\unit[260]{\mu m}$ between the corresponding ions.

\subsubsection{Second sequence: transforming a 1D cluster state into a 2D cluster state}
\label{sec:Transform-1d-2d} Now that a linear cluster state has been
prepared during the first sequence of operations, qubits 1 and 4, 3
and 6, and 5 and 8 need to be entangled in order to
prepare a two-dimensional cluster state during the second sequence.
This can be achieved in three steps. In the first step of the second
sequence during time interval $t_2 < t< t_3$, qubits 1 through 4 are
stored in a common trap, e.g., with trap frequency $\omega =
2\pi\times \unit[200]{kHz}$, while all other ions occupy single,
individual traps with trap distances of $\unit[260]{\mu m}$. Now,
applying a magnetic-field gradient of $\unit[100]{T/m}$ results in a
coupling of spins 1 and 4 of $J_{14} = \unit[1.24]{kHz}$ with time-
evolution operator
 \be
 U(t_{14}) = \exp(-\frac{i}{2} \sum_{n,m=1; n \le m}^4 J_{n,m} t_{14} \sigma_{z,n}\sigma_{z,m}).
 \ee
Other couplings involving qubits 1 to 4 can be canceled by selective
recoupling pulse sequences \cite{Leung00}. All couplings
regarding qubits 5 to 8 can be neglected during this time interval,
since those are approximately $\unit[0.5]{Hz}$. The entire time
evolution for coupling ions 1 and 4 including pulses to eliminate
couplings involving qubits 2 and 3 is thus given by
 \be
\begin{split}
U_{t_3} = &U(t_{14}/4) \sigma_{x,2} U(t_{14}/4) \sigma_{x,2} \times
\\
&
\times
\sigma_{x,3} U(t_{14}/4) \sigma_{x,2} U(t_{14}/4) \sigma_{x,2} \sigma_{x,3},
\end{split}
 \ee
where $t_{14} = \frac{\pi}{2 J_{14}}$ and $\sigma_{x,k}$ denotes the
usual Pauli $X$ operator acting on the Hilbert space of qubit $k$.
To see that $U_{t_3}$ implements only the coupling between qubits
$1$ and $4$, it is convenient to think of the two rows as individual
sequences. In each of the two sequences, all couplings involving
qubit $2$ are eliminated, because applying $\sigma_{x,2}$ before and
after $U(t_{14}/4)$ adds a minus sign to every $\sigma_{z,2}$ in
this time-evolution due to the commutation relation of the Pauli
matrices. Analogously, the $\sigma_{x,3}$ at the beginning and end
of the second sequence affects nothing but adding a minus sign to
all couplings involving qubit 3. In total, all couplings to qubits
$2$ and $3$ differ in their signs for exactly half of the time, and
thus they are eliminated. On the other hand, the coupling of qubit
$1$ to $4$ has the same sign during the entire sequence, such that
$U_{t_3} = \exp(-i \frac{\pi}{4} \sigma_{z,1} \sigma_{z,4})$.

At the end of time interval $t_{14}$, the magnetic-field gradient is
switched off, and the same procedure is repeated to entangle ions 3
and 6, respectively 5 and 8. Then the ions are arranged in a
configuration, such that ions 3 to 6 occupy a common potential well
and all other ions are located in separate trap potentials.
Switching on the magnetic-field gradient again results in \be
 U(t_{36}) = \exp(-\frac{i}{2} \sum_{n,m=3; n \le m}^6 J_{n,m} t_{36} \sigma_{z,n}\sigma_{z,m}),
\ee and the time evolution including recoupling pulses reads \be
\begin{split}
U_{t_4} = & U(t_{36}/4) \sigma_{x,4} U(t_{36}/4) \sigma_{x,4} \times
\\
&
\times
\sigma_{x,5} U(t_{36}/4) \sigma_{x,4} U(t_{36}/4) \sigma_{x,4} \sigma_{x,5},
\end{split}
\ee
with $t_{36} = \frac{\pi}{2 J_{36}}$. Repeating this procedure for entangling
ions 5 and 8, one obtains the following time evolution:
\be
\begin{split}
U_{t_5} = & U(t_{58}/4) \sigma_{x,6} U(t_{58}/4) \sigma_{x,6} \times
\\
&
\times
\sigma_{x,7} U(t_{58}/4) \sigma_{x,6} U(t_{58}/4) \sigma_{x,6} \sigma_{x,7},
\end{split}
\ee
where $t_{58} = \frac{\pi}{2 J_{58}}$ and
\be
 U(t_{58}) = \exp(-\frac{i}{2} \sum_{n,m=5; n \le m}^8 J_{n,m} t_{58} \sigma_{z,n}\sigma_{z,m}).
 \ee
The result of this procedure is the cluster state based on the graph
shown in Fig. \ref{fig:cluster8}:
 \be
 \begin{split}
  |\Psi\rangle = & U_{t_5} U_{t_4} U_{t_3} U_{t_2} U_{t_1} |+\rangle^{\otimes^8}
  \\
  = & \exp(-i\frac{\pi}{4} (\sigma_{z,1}\sigma_{z,4} +\sigma_{z,3}\sigma_{z,6}+
  \\
  &
  + \sigma_{z,5}\sigma_{z,8}+\sum_{i=1}^7 \sigma_{z,i}\sigma_{z,i+1}) |+\rangle^{\otimes^8}.
  \end{split}
 \ee
This scheme can be scaled to any $n \times 2$ cluster, which may
then serve to simulate a $n\times m$ cluster.

\subsubsection{Summary of transport scheme for generating 2D cluster states}

\begin{table*}
\begin{tabular}{|c|c|}
\hline Step & Operation
\\
\hline 1 & Transporting ions 1/2, 3/4, 5/6, 7/8  in common trap
potentials
\\
\cline{2-2} & Entangling ions 1/2, 3/4, 5/6, 7/8
\\
\hline 2 & Transporting ions 2/3, 4/5, 6/7 in common trap potentials
\\
\cline{2-2} & Entangling of ions 2/3, 4/5, 6/7
\\
\hline 3 & Recombination of ions 1 - 4 and 5 - 8
\\
\cline{2-2} & Entangling ions 1 and 4 using selective recoupling
\\
\hline 4 & Recombination of ions 3 - 6
\\
\cline{2-2} &  Entangling ions 3 and 6 using selective recoupling
\\
\hline 5 & Recombination of ions 5 - 8
\\
\cline{2-2} & Entangling ions 5 and 8 via selective recoupling
\\
\hline
\end{tabular}
\caption{Summary of transport scheme for preparing a $n \times 2$
cluster state in a segmented ion trap with magnetic-field gradient
of $\unit[100]{T/m}$. Steps 1 through 2 serve to create a linear
cluster state, while steps 3 through 5 create additional third
neighbor couplings that turn the one-dimensional cluster into a
two-dimensional cluster. Here, the coupling constants $J_{i,i+1} =
\unit[3.0]{kHz}$ for preparation of the linear cluster state which
yields the gate times $t_j=\frac{\pi}{2J_{i,i+1}} =
\unit[0.52]{ms}$, $j=1,2$. In the case of 4 ions sharing a common
potential well $J_{1,4}, J_{3,6}, J_{5,8}$ are given by
$\unit[1.2]{kHz}$ and we have $t_j=\frac{\pi}{2J_{i,i+1}} =
\unit[1.3]{ms}$, $j=3,4,5$. The time scale for adiabatically
transporting ions is $t_T \gg 2\pi / \nu_1 \approx$
\unit[5]{$\mu$s}. The time scale for adiabatically turning on and
off the magnetic field is $t_B \gg 2\pi / \Delta$. When using
\ybuion ions the relevant $\Delta$ indicates the Zeeman splitting of
the ($S_{1/2}$, $F=1$) state and $2\pi/\Delta$ is typically of
order \unit[0.1]{$\mu$s} }
 \label{tab:SchemeSummary}
\end{table*}

A summary of the scheme that makes use of transport of ions over
small distances is provided in Table \ref{tab:SchemeSummary}. If we
restrict the scheme to adiabatic ion transport, the required time
$t_T$ for transport must obey $t_T
>> 2\pi / \nu_1$.
For $\nu_1 \approx 2\pi \times \unit[200]{kHz}$, we estimate $t_T =
\unit[50]{\mu s}$ which is still more than 1 order of magnitude
less than the time required for entangling the qubits. The ions need
only be linearly transported over distances of the order of the size
of two electrode segments; in our concrete example considered above
this amounts to a distance of around \unit[200]{$\mu$m}.

Adiabatic switching of the magnetic-field gradient would require a
time scale of order $t_{B} >> \frac{2\pi}{\omega}$, if only the
qubit states $|0\rangle$ and $|1\rangle$ were present. However,
usually other nearby ionic states have to be included in these
considerations. For example, in the case of \ybuion ions and
choosing $|0\rangle\equiv |S_{1/2}, F=0\rangle$ and
$|1\rangle\equiv |S_{1/2}, F=1, m_F=1 \rangle$ one would
require $t_{B} >> \frac{2\pi}{\Delta}$ to avoid transitions between
Zeeman states, where $\Delta$ indicates the Zeeman splitting of
the ($S_{1/2}$, $F=1$) state. In order to avoid zero crossings
of the states, while the magnetic-field gradient is changed, a
constant offset field is used, e.g., such that $\Delta = \unit[2\pi
10]{MHz}$ without gradient, so that we estimate $t_{B} \approx
\unit[1]{\mu s}$.

During this switching process the qubits' phases are affected due to
the spin-spin coupling and the evolution of the Zeeman state.
Applying the decoupling scheme described in \cite{Leung00} with
appropriate time intervals removes changes in the qubits' phases due
to this Zeeman evolution together with the undesired spin-spin
couplings as described above.

\section{Conclusion}

We propose methods to generate cluster states of trapped ions
confined in state-of-the-art segmented linear ion traps by
engineering their spin-spin coupling constants. Based on the idea of
simulating a $n\times m$ cluster by a $n \times 2$ cluster within
the one-way model of quantum computing, we examined in Sec.
\ref{sec:ions-confined} a method previously suggested
\cite{McHugh05} to prepare $n\times 2$ clusters, and a novel idea
based on creating individual potential wells for each ion. In
addition, the superposition of harmonic potentials is discussed in
order to engineer $J$ couplings fulfilling suitable periodicity
relations to create small cluster states in one time-evolution step.
In order to achieve sizeable coupling constants (in the kilohertz range),
all these methods require control over local electrostatic
potentials with a spatial resolution of the order \unit[10]{$\mu$m}
-- a typical interion distance in usual Paul traps. Therefore, it
is of interest to investigate how suitable a trap with larger
electrode structures is for generating 2D cluster states.

In Sec. \ref{sec:including-ion-transport} a scheme for preparing
$n\times 2$ clusters is described for which larger trap electrode
structures (of the order \unit[100]{$\mu$m}) are sufficient. Here,
$n\times 2$ clusters are prepared by first creating a linear cluster
and subsequently enabling third-neighbor couplings. We showed that the
generation of the linear cluster state can be accomplished with
modern segmented ion traps by locating pairs of two ions in common
harmonic oscillators, thus resulting in uniform NN couplings. After
separating the ions and subsequently merging them with the other
nearest-neighbors, the second half of NN couplings are realized. The
required third-neighbor couplings are achieved via selective
recoupling techniques.

Common to all schemes described in this paper is that entanglement
is achieved solely by controlling dc voltages and currents: no
coherent interaction between laser light and trapped ions is
required. Refocusing pulses applied to individual ions consist of
radio-frequency or microwave radiation depending on the choice of
qubit \cite{Mintert2001,Wunderlich02,Wunderlich2003,Johanning2008}.

The method described in Sec. \ref{sec:including-ion-transport} is explicitly worked
out for the generation of a eight-qubit 2D cluster state. This scheme
is also applicable to the generation of $n\times 2$ cluster states
where $n=k\times 4$, $k = 2,3,4, ...$. The recipe would then be to
create in parallel $k$ 2D cluster states of size $4\times 2$. Then
in, one additional step, ions at the edges of neighboring $4\times
2$ clusters are entangled. For example, generating a $8\times 2$
cluster proceeds as follows: first create two 2D cluster states of
size $4\times 2$, then by combining ions number 7, 8, 9, and 10 in
one potential well, simultaneously entangle ion pairs 8/9 and 7/10 to
complete a 16-qubit 2D cluster state.

The physical arrangement of a 2D cluster state of size $n\times 2$
is not required to be a linear ion string as was assumed so far. For
instance, eight ions could reside in one area of a large 2D trap array
\cite{Kielpinski2002} and communication between different areas, and
thus entanglement, could be achieved by shuttling only the ions at
the ends of each ion string.

The spin-spin coupling that is used here does not require cooling of
the ion string to its motional ground state. Detailed calculations
show that cooling to the Doppler limit is sufficient for suppressing
unwanted effects of thermal motion. This is beyond the scope of this
paper and will be subject of a separate publication.

\section*{Acknowledgements}
We acknowledge financial support by the STREP Microtrap funded by
the European Union, the European Union IP QAP, by the Deutsche
Forschungsgemeinschaft, and by secunet AG.

\end{document}